\documentclass{article}
\usepackage{spconf,amsmath,graphicx}

\usepackage{bm}
\usepackage{booktabs}
\usepackage{array}
\newcolumntype{M}[1]{>{\centering\arraybackslash}m{#1}}
\usepackage{arydshln}
\DeclareMathOperator*{\argmin}{\arg\!\min}

\title{Robust End-to-End Diarization with Domain Adaptive Training and Multi-Task Learning}
\name{Ivan Fung, Lahiru Samarakoon, Samuel J. Broughton}
\address{Fano Labs, Hong Kong}
\copyrightnotice{979-8-3503-0689-7/23/\$31.00~\copyright2023 IEEE}
\begin{document}
\ninept
\maketitle
\begin{abstract}
Due to the scarcity of publicly available diarization data, the model performance can be improved by training a single model with data from different domains. In this work, we propose to incorporate domain information to train a single end-to-end diarization model for multiple domains. First, we employ domain adaptive training with parameter-efficient adapters for on-the-fly model reconfiguration. Second, we introduce an auxiliary domain classification task to make the diarization model more domain-aware. For seen domains, the combination of our proposed methods reduces the absolute DER from 17.66\% to 16.59\% when compared with the baseline. During inference, adapters from ground-truth domains are not available for unseen domains. We demonstrate our model exhibits a stronger generalizability to unseen domains when adapters are removed. For two unseen domains, this improves the DER performance from 39.91\% to 23.09\% and 25.32\% to 18.76\% over the baseline, respectively.

\end{abstract}
\noindent\textbf{Index Terms}: domain adaptive training, multi-task learning, adapter, domain classification, end-to-end speaker diarization

\section{Introduction}

Speaker diarization refers to the task of ``who spoken when''. In traditional approaches towards diarization, an audio recording is first split into segments with a speech activity detection model. Segments are then used for extracting speaker embeddings \cite{xvector,dvector}. Based on the speaker embeddings, segments are partitioned into speaker clusters by using clustering algorithms \cite{cluster1, cluster2, cluster3}. There are two major disadvantages for the traditional approach. First, it is not able to deal with overlapping speech from multiple speakers, which is important for practical use. Second, it is difficult to optimize and maintain multiple components for only one diarization task.

End-to-end diarization (EEND) models achieve the state-of-the-art performance by estimating speech activities from multiple speakers using a single unified model \cite{eend, sa_eend}. However, EEND models have a limitation of only being able to output a fixed number of speakers. EEND models with encoder-decoder based attractors (EEND-EDA) are used to support an unknown number of speakers, which achieve superior DER performance \cite{eend_eda}. EEND-EDA models have been used as the base architecture in a number of recent research \cite{online1, online2, eend_eda_aam, sam}.

When the amount of training data in each domain is scarce, the model performance can be improved by training a single model with data from different domains. This is because it improves the generalization capability of the model. Domain adaptive training (DAT) allows one single model to be optimized to multiple domains. When adapters are used in DAT, the model can be reconfigured on-the-fly to different seen domains \cite{adapter}. Adapters are domain-specific which only contain a small fraction of model parameters. The majority of parameters in the model remain domain-agnostic for common representation learning across different domains. Under the area of computer vision, DAT with adapters has been shown to achieve the state-of-the-art performance \cite{adapter}. In recent years, the utilization of DAT with adapters has also been successfully applied in a great variety of areas to improve model performance in different domains. For instance, it has been adopted in multilingual automatic speech recognition (ASR) \cite{adapter_multi_streaming}. With a similar motivation to adapters, DAT with domain-specific layers have also been exploited in ASR \cite{configurable}. However, the use of DAT with adapters has not been explored and examined in speaker diarization before.

Multi-task learning (MTL) improves the model generalization performance to multiple tasks through solving these tasks simultaneously \cite{mtl}. These models first learn a common representation for different tasks with shared parameters. Task-specific parameters are then used to learn the transformation from the original representation with respect to each task. Depending on the number of tasks present in the MTL framework, the corresponding number of objectives are used during model training. The technique of multi-task learning has been successfully employed in a wide range of areas over the last few decades. Under the area of automatic speech recognition, MTL with speech enhancement \cite{mtl_asr1}, and different target prediction \cite{mtl_asr2} have been explored to improve model performance. In speaker verification, MTL with word prediction \cite{mtl_word}, phonetic prediction \cite{mtl_phone}, and speaker attribute prediction \cite{mtl_speaker} have been used to learn robust speaker representation for superior model performance. In speaker diarization, using MTL with ASR \cite{mtl_diar_asr}, speech activity detection and overlap detection \cite{mtl_diar_vad}, and speech separation \cite{mtl_diar_ss} have also been shown effective. Under the MTL framework, an additional domain classification task can be used to improve the model capability of domain prediction. When training a single model with data combined from multiple domains, the domain classification task makes the diarization model more aware to different domains. However, using MTL with a such auxiliary domain classification task has not been studied in speaker diarization before.

In this work, we propose to leverage domain information to train a single end-to-end diarization model for multiple domains. We employ two different techniques, namely domain adaptive training with adapters, and multi-task learning with domain classification. First, domain adaptive training with adapters enables domain-specific on-the-fly model reconfiguration in a parameter-efficient way. This allows a single diarization model to be optimized to multiple domains. Second, multi-task learning with domain classification improves the model capability of domain prediction, which makes the diarization model more domain-aware. One drawback of using adapters is that we need to know the domain information in advance. We investigate multiple methods to overcome this limitation when performing diarization in unseen domains. We demonstrate our model achieves a stronger generalizability to unseen domains over a strong multi-domain baseline. To the best of our knowledge, this is the first work to utilize domain information to train a single end-to-end diarization model for multiple domains with the aforementioned techniques.

\newpage

\section{Background}

In this section, we review the Conformer encoder and the end-to-end diarization model used in this work.

\subsection{Conformer encoder}

Transformer uses a self-attention module to capture global dependency \cite{transformer}. Conformer is a variant of Transformer that is proposed to capture local dependency with an extra convolution module in each encoder block, which has been shown to achieve the state-of-the-art performance in ASR \cite{conformer}. The equations for each Conformer encoder block are formulated as follows:
\begin{align}
\label{eq:conformer}
	\tilde{X} &= X + \frac{1}{2} \text{FF}(X) \\
    X' &= \tilde{X} + \text{SA}(\tilde{X}) \\
    X'' &= X' + \text{CNN}(X') \\
    X''' &= \text{LN}(X'' + \frac{1}{2} \text{FF}(X'')),
\end{align}
where $X$ and $X'''$ are the input and output of each encoder block, FF is the feed forward module, SA is the self-attention module, CNN is the convolution module and LN is layer normalization.

Conformer has also been shown to achieve the state-of-the-art performance in end-to-end diarization \cite{conformer2,conformer3,conformer4}. In this work, we also adopt Conformer as the backbone of our diarization model.

\subsection{End-to-end diarization}

In EEND, the diarization loss $L_{diar}$ with permutation-invariant training (PIT) is used \cite{eend, sa_eend}:
\begin{align}
\label{eq:eend}
	L_{diar} = \frac{1}{TS} \argmin_{\phi \in \text{perm}(1, \ldots, S)} \sum_{t=1}^T H(\bm{y}_t^\phi, \bm{\hat{y}}_t),
\end{align}
where $T$ denotes the number of frames and $S$ is the number of speakers. Here, $H(\bm{y}_t^\phi, \bm{\hat{y}}_t)$ is the binary cross entropy between permuted speaker labels $\bm{y}_t^\phi$ and ground-truth speaker labels $\bm{\hat{y}}_t$.

The addition of an encoder-decoder based attractor calculation (EDA) module enables the diarization model to handle a flexible number of speakers \cite{eend_eda}. The EDA module generates speaker-wise attractors iteratively, which are used to estimate the total number of speakers in the audio. The attractor existence loss $L_{attr}$ can be formulated as below:
\begin{align}
\label{eq:eda}
    \bm{h}_0, \bm{c}_0 &= \text{EDA}_{enc} (\bm{e}_1, \ldots, \bm{e}_T) \\
    \bm{h}_s, \bm{c}_s, \bm{a}_s &= \text{EDA}_{dec} (\bm{h}_{s-1}, \bm{c}_{s-1}, \bm{0}) \\
    p_s &= \sigma(\bm{w}^\top \bm{a}_s + b) \\
    L_{attr} &= \frac{1}{S+1} H(\bm{l}, \bm{p}),
\end{align}
where $(\bm{e_t})_{t=1}^T$ denotes the embedding output sequence from EEND. Here, $\bm{h_s}$ and $\bm{c_s}$ are the hidden state and cell state at time step $s$ in the decoder. $\bm{a}_s$ denotes the attractor for speaker $s$. $\bm{w}$ and $b$ are trainable parameters. $H(\bm{l}, \bm{p})$ is the binary cross entropy between ground-truth speaker existence indicator variable $\bm{l}$ and attractor existence probability $\bm{p}$.

When the EDA component is used, $\bm{y}_t^\phi$ in Equation \ref{eq:eend} is computed based on the attractors instead:
\begin{align}
\label{eq:eday}
    y_{t,s}^\phi = \sigma(\bm{a}_s^\top\bm{e}_t),
\end{align}

The final training objective $L_{final}$ is a combination of $L_{diar}$ and $L_{attr}$, which can be formulated below:
\begin{align}
\label{eq:loss1}
	L_{final} = L_{diar} + \alpha L_{attr},
 \end{align}
where $\alpha$ is the weighting hyper-parameter.

\begin{figure}[htb]
  \centerline{\includegraphics[width=7.25cm]{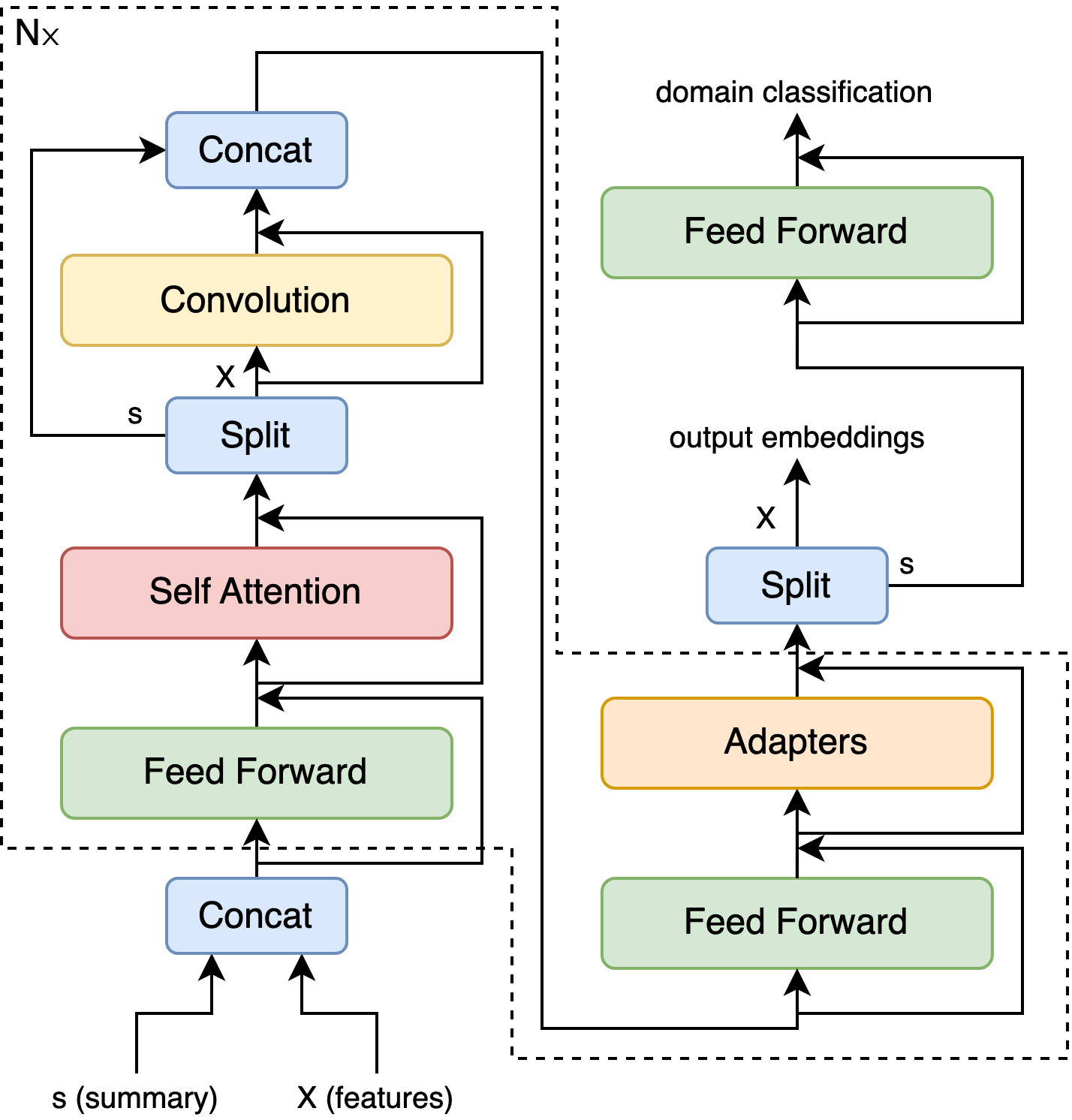}} 
  \caption{The architecture of Conformer encoder contains (1) domain-adaptive training with adapters in all encoder blocks, and (2) multi-task learning with an auxiliary domain classification task using the conversational summary vector.}
  \label{fig:conformer}
\end{figure}

\section{Proposed Methods}

In this section, we propose to leverage domain information to train a single end-to-end diarization model for multiple domains. First, we discuss domain-adaptive training with adapters. Second, we examine multi-task learning with an auxiliary domain classification task. Lastly, we explore the removal of adapters for unseen domains. The architecture of Conformer encoder with our proposed methods is illustrated in Figure \ref{fig:conformer}.

\subsection{Domain-adaptive training with adapters}

\begin{figure}[hbt]
  \centerline{\includegraphics[width=3cm]{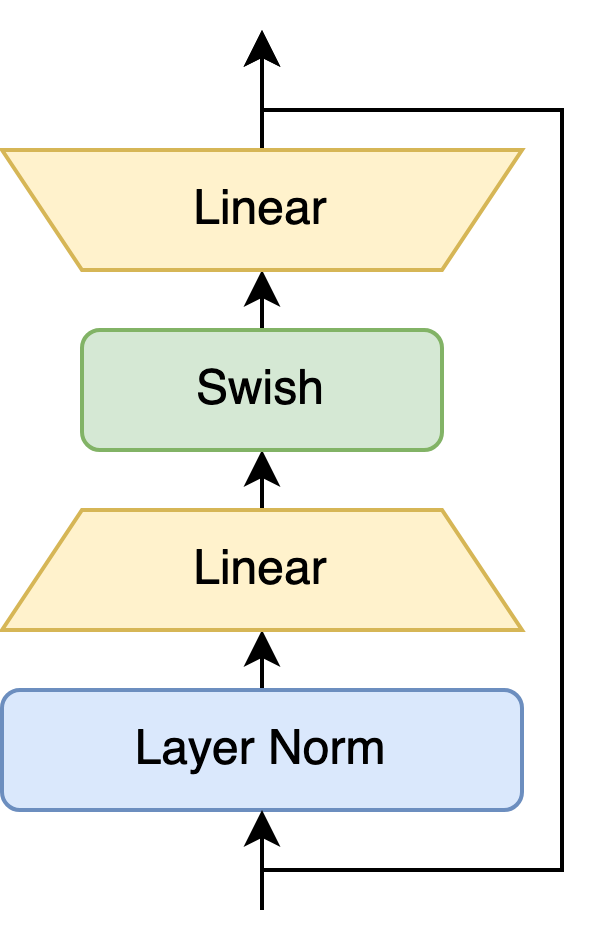}}  
  \caption{Architecture of a single adapter module. The input is first projected from the original to the bottleneck dimension with Swish activation. It is then projected back to the original dimension.}
  \label{fig:adapter}
\end{figure}

Domain-adaptive training with adapters optimizes a single model to multiple domains. This allows on-the-fly reconfiguration of the model to different domains of interest in a parameter-efficient way \cite{adapter}. Adapters are domain-specific and only composed of a small fraction of model parameters. The majority of parameters in the model remain domain-agnostic for common representation learning across different domains.

As depicted in Figure \ref{fig:adapter}, each adapter module consists of a layer normalization and two linear projections with an in-between Swish activation on the bottleneck features. The adapters from different domains are appended after each encoder block. This facilitates an adapter to learn the domain-wise adjustment with respect to a specific domain. The equation of an adapter can be formulated below:
\begin{equation}
    X'''' = X''' + \text{Linear}^{(d)}_2(\text{Swish}(\text{Linear}^{(d)}_1(\text{LN}^{(d)}(X'''))))
\end{equation}
where $d$ denotes the domain. Here, $\text{LN}^{(d)}$, $\text{Linear}^{(d)}_1$ and $\text{Linear}^{(d)}_2$ are the layer normalization, linear projection from original to bottleneck dimension, and linear projection back to the original dimension respectively for each domain $d$.

For any seen domain, the adapter from the ground-truth domain is used during inference. For unseen domains, adapters are removed during inference, which will be discussed in Section \ref{ssec:removal}.

\subsection{Multi-task learning with domain classification}
\label{ssec:mtl}

Multi-task learning (MTL) improves model generalization performance to multiple task through solving these task simultaneously \cite{mtl}. The model capability of domain prediction can be improved by an additional domain classification task. When training a single model with data combined from multiple domains, this makes the diarization model more aware to different domains.

We explore two areas in which the domain classification objective is used. The first area investigates inputting the final EDA encoder states ($\bm{h}_0$, $\bm{c}_0$) to domain classification. It is natural because these states already contain the aggregated conversation-level information. The second makes use of the conversational summary representation for EEND-EDA from a recent research \cite{sam}. This involves a conversational summary vector ($\bm{s}$) that is inspired by the classification token in  BERT \cite{bert}. It is a learnable input embedding that can be concatenated with the audio feature as model input. The conversational summary representation maintains a global representation through bypassing the convolution module in the encoder. The input to domain classification is thus one of the below vectors:
\begin{align}
\bm{v} = \bm{h}_0 \oplus \bm{c}_0, \quad \text{or} \quad \bm{v} = \bm{s},
\end{align}
where $\oplus$ is the concatenation operation.

The auxiliary domain classification task with the cross-entropy loss can be formulated below:
\begin{align}
\label{eq:domain}
    p(y^{(d)} | X) &= \text{Softmax} (\bm{v} + \text{FF}_{dc}(\bm{v})) \\
	L_{domain} &= -\log p(y^{(d)} | X),
\end{align}
where $y^{(d)}$ is ground-truth domain label and $\text{FF}_{dc}$ is the feed forward module for the auxiliary domain classification task. The hidden dimension of this feed forward module is the same as the input.

When $\bm{v} = \bm{s}$ is used, this vector can also bypass the adapters in the encoder. This can make the auxiliary domain classification task more difficult because it has not been adapted to a specific domain before being used for domain classification. This can be done by a similar split and concatenation mechanism used in the convolution module above. The effect of bypassing this vector in adapters will be investigated in Section \ref{ssec:bypass}.

The final training objective is updated to $L'_{final}$, which is a combination of all three losses:
\begin{align}
\label{eq:loss2}
	L'_{final} &= L_{diar} + \alpha L_{attr} + \beta L_{domain},
 \end{align}
where $\beta$ is the weighting hyper-parameter.

The auxiliary domain classification task is not required during inference. It can be removed to save computational cost.

\subsection{Removal of adapters for unseen domains}
\label{ssec:removal}

For seen domains, adapters from the ground-truth domains can be utilized during inference for optimized performance. For unseen domains, all adapters are removed during inference because adapters from the ground-truth domains are not available. In this case, only domain-agnostic components in the model are utilized during inference. However, this can possibly harm the model performance due to training and testing mismatch. In Section \ref{ssec:generalizbility}, we study the effect of using adapter dropout, which removes adapters during training based on a certain probability.

\begin{table}[htb]
\caption{Summary of publicly available datasets used in finetuning and inference stages.}
\label{tab:datasets}
\centering
\resizebox{0.975\linewidth}{!}{%
    \begin{tabular}{ccM{1.5cm}M{1.5cm}}
    \toprule
    \textbf{Stage} & \textbf{Dataset} & \textbf{No. of Speakers} & \textbf{No. of Mixtures} \\ \midrule
    Finetuning 
        & \begin{tabular}[c]{@{}c@{}}
            DIHARD III Dev (Train)\\
            VoxConverse Dev\\
            MagicData-RAMC Train\\
            AMI MIX Train\\
            AMI SDM1 Train
        \end{tabular}
        & \begin{tabular}[c]{@{}c@{}}
            1 - 10\\
            1 - 20\\
            2\\
            3 - 5\\
            3 - 5
        \end{tabular}
        & \begin{tabular}[c]{@{}c@{}}
            203\\
            216\\
            289\\
            136\\
            135
        \end{tabular} \\
        \midrule
    Inference
        & \begin{tabular}[c]{@{}c@{}}
            DIHARD III Eval\\
            VoxConverse Test\\
            MagicData-RAMC Test\\
            AMI MIX Eval\\
            AMI SDM1 Eval\\
            AliMeeting Far Test\\
            AliMeeting Near Test\\
        \end{tabular}
        & \begin{tabular}[c]{@{}c@{}}
            1 - 9\\
            1 - 21\\
            2\\
            3 - 4\\
            3 - 4\\
            2 - 4\\
            2 - 4\\
        \end{tabular}
        & \begin{tabular}[c]{@{}c@{}}
            259\\
            232\\
            43\\
            16\\
            16\\
            20\\
            20\\
        \end{tabular} \\
        \bottomrule
    \end{tabular}
}
\end{table}

\section{Experiments}

\subsection{Datasets}
\label{datasets}

The LibriSpeech \cite{librispeech} corpus is used for pre-training. A total number of 400,000 mixtures are simulated following the same protocols as explained in \cite{eend_eda_aam}. Five publicly available datasets are used for finetuning, namely DIHARD III (DH3) \cite{dihard3}, VoxConverse (VC) \cite{voxconverse}, MagicData-RAMC (MD) \cite{magicdata}, AMI MIX (AM) \cite{ami} and AMI SDM1 (AS) \cite{ami}, as described in \cite{sam}. This consists of 979 mixtures in total. The details for each training and test set have been summarized in Table \ref{tab:datasets}. DIHARD III is collected from diverse environments which include recordings from audiobooks, interviews, courtroom, telephone, meetings, restaurants and video websites. VoxConverse is extracted from YouTube videos with challenging background conditions. MagicData-RAMC contains conversational speech recorded  by mobile phones. AMI MIX and AMI SDM1 consist of meetings recorded by close-talking and far-field microphones respectively. Here, we treat each dataset as one single domain. The 11-domain DIHARD III dataset is not used to evaluate our proposed methods due to data scarcity in each of its domains, which can lead to unstable diarization results. All finetuning data is augmented with the MUSAN corpus \cite{musan}, which includes babble, music, noise and noise\_bg.

During inference, we treat the above 5 datasets as 5 different seen domains. In addition, we make use of AliMeeting Far (AF) \cite{ali} and AliMeeting Near (AN) \cite{ali} to evaluate the diarization performance for unseen domains. AliMeeting Far and AliMeeting Near contain meetings recorded by far-field microphone arrays and near-field microphone headsets respectively.

\subsection{Experimental setup}

We follow the training procedure described in \cite{eend_eda_aam} as the default, which consists of a Conformer with 4 encoder blocks as the backbone. In each of the blocks, a hidden size of 256 units with 4 attention heads is used in self-attention. A hidden size of 1,024 is used in feed forward module. Convolution sub-sampling and up-sampling modules are exploited with the same settings in \cite{eend_eda_aam}. No positional encodings are used. Frame-wise augmentation using random frame shuffling before the EDA component is used \cite{sam}. Gradient cutting for attractor existence loss is applied before the EDA component to avoid back propagation to the encoder. During pre-training, the Noam scheduler \cite{transformer} with 100,000 warm-up steps are used. A batch size of 64 and a learning rate of $5e^{-5}$ with Adam optimizer are used throughout all training stages. The weighting hyper-parameter $\alpha$ for speaker existence loss is set to 1.0. The additive angular margin loss is used with $m$ of 0.35 and $\gamma$ of 10, as recommended in \cite{eend_eda_aam}. A maximum number of 4 dominant speakers are used during training due to the computational complexity of PIT in diarization loss \cite{eend_eda}.

The model is first pre-trained with LibriSpeech mixtures of 2 speakers for 100 epochs. With weights initialised from checkpoint averaging of the last 10 epochs, it is then pre-trained with mixtures of all speakers for another 25 epochs. For each individual experiment, the pre-trained model is finetuned for another 500 epochs. Each training sample is randomly cropped from the original audio into a length of 50 seconds, which is equivalent to 5,000 frames. The input feature is extracted with 23 dimensional log Mel-filterbanks using a window size of 25 ms and a hop size of 10 ms.

All experimental results are evaluated with a diarization threshold of 0.5, an attractor existence threshold of 0.5, a median filtering of 11 frames and a collar of 0.0. During inference, the model is averaged with checkpoints from the 10-best epochs based on validation DER. To stay consistent to the training configuration, each evaluation set is reported up to 4 speakers only, unless otherwise specified.

\begin{table}[htb]
  \caption{DER performance of domain-specific (B1) and multi-domain (B2) baseline models.}
  \label{tab:baseline}
  \centering
  \resizebox{0.875\linewidth}{!}{
  \begin{tabular}{ccccccc}
    \toprule
    Setup & DH3 & VC & MD & AM & AS & Overall \\
    \midrule
    B1 & \textbf{13.48} & 17.18 & 19.74 & 21.36 & 34.15 & 19.24 \\
    B2 & 14.25 & \textbf{12.87} & \textbf{15.75} & \textbf{20.23} & \textbf{33.71} & \textbf{17.66} \\
    \bottomrule
  \end{tabular}}
\end{table}

\section{Results}
In Table \ref{tab:baseline}, we provide the DER performance for the baseline models. Each domain-specific model in B1 is trained with a single domain only, while the multi-domain model (B2) is trained with all five seen domains. It can be observed that B2 has better performance in all domains except for DH3, and improves the overall absolute DER from 19.24\% to 17.66\%. This contributes to a relative improvement of 8.2\% in comparison with B1. Since the amount of training data in B1 is scarce, these results suggest that the domain-specific models in B1 are not trained with sufficient data. Next, we leverage domain information to train a single diarization model for multiple domains.

\begin{table}[htb]
  \caption{Effect of using various weight ($\beta$) in the domain classification loss. LSTM encoder states ($\bm{v} = \bm{h}_0 \oplus \bm{c}_0$) are used as the input to the domain classification task.}
  \label{tab:mtl_lstm}
  \centering
  \resizebox{\linewidth}{!}{
  \begin{tabular}{cccccccc}
    \toprule
    Setup & Weight ($\beta$) & DH3 & VC & MD & AM & AS & Overall \\
    \midrule
    B2 & - & 14.25 & \textbf{12.87} & 15.75 & \textbf{20.23} & 33.71 & \textbf{17.66} \\
    M1 & 1.0 & \textbf{14.06} & 13.62 & 16.49 & 23.15 & 34.04 & 18.30 \\
    M2 & 2.0 & 14.46 & 14.72 & 15.01 & 22.29 & \textbf{33.42} & 18.04 \\
    M3 & 3.0 & 14.32 & 15.05 & \textbf{14.70} & 21.89 & 34.58 & 18.06 \\
    M4 & 4.0 & 14.44 & 15.17 & 14.96 & 29.08 & 36.71 & 19.36 \\
    \bottomrule
  \end{tabular}}
\end{table}

Here, we study the effectiveness of MTL with an auxiliary domain classification task. In Table \ref{tab:mtl_lstm}, a slight DER degradation to B2 can be observed when the last LSTM encoder states ($\bm{h}_0$, $\bm{c}_0$) from EDA are used for domain classification. We hypothesize that these states are optimized to capture speaker characteristics for attractor existence estimation. It is possible that the auxiliary domain classification task confuses the speaker representation learning in attractors.

\begin{table}[htb]
  \caption{Effect of using various weight ($\beta$) in the domain classification loss. Conversational summary vector ($\bm{v} = \bm{s}$) is used as the input to the domain classification task.}
  \label{tab:mtl_summary}
  \centering
  \resizebox{\linewidth}{!}{
  \begin{tabular}{cccccccc}
    \toprule
    Setup & Weight ($\beta$) & DH3 & VC & MD & AM & AS & Overall \\
    \midrule
    B2 & - & 14.25 & 12.87 & 15.75 & 20.23 & 33.71 & 17.66 \\
    M5 & 1.0 & \textbf{13.79} & 13.86 & 17.98 & 18.88 & 33.31 & 17.99 \\
    M6 & 2.0 & 14.19 & \textbf{12.62} & \textbf{14.33} & \textbf{18.87} & \textbf{33.12} & \textbf{16.99} \\
    M7 & 3.0 & 14.17 & 14.44 & 17.30 & 23.32 & 37.42 & 19.12 \\
    M8 & 4.0 & 14.05 & 13.86 & 20.14 & 19.51 & 36.18 & 19.07 \\
    \bottomrule
  \end{tabular}}
\end{table}

In Table \ref{tab:mtl_summary}, the conversational summary vector is utilized for the auxiliary domain classification task instead. When the weight in the domain classification loss is 2.0, the DER performance is improved consistently over all five domains. The absolute DER performance of M6 improves from 17.66\% to 16.99\% when compared with B2, which contributes to a relative improvement of 3.8\%. This demonstrates the addition of an auxiliary domain classification task can improve diarization performance. A possible reason is that the diarization model is more aware to different domains.

\begin{table*}[htb]
  \caption{Effect of domain adaptive training (DAT) with adapters in various encoder block(s) using different bottleneck size.}
  \label{tab:dat}
  \centering
  \resizebox{0.85\linewidth}{!}{
  \begin{tabular}{cccccccccc}
    \toprule
    Setup & No. of Parameters & Encoder Block(s) & Bottleneck Size & DH3 & VC & MD & AM & AS & Overall \\
    \midrule
    B2 & 8.10M & - & - & 14.25 & 12.87 & 15.75 & 20.23 & 33.71 & 17.66 \\
    A1 & 8.28M & all & 16 & 12.94 & 12.79 & 18.89 & 19.57 & \textbf{30.49} & 17.50 \\
    A2 & 8.45M & all & 32 & 12.94 & \textbf{11.73} & 16.51 & 19.10 & 30.99 & \textbf{16.74} \\
    A3 & 8.78M & all & 64 & 13.17 & 12.44 & 18.98 & 19.77 & 33.01 & 17.90 \\
    A4 & 9.43M & all & 128 & 12.95 & 14.67 & 17.93 & \textbf{17.90} & 33.49 & 17.70 \\
    \hdashline
    \rule{0pt}{0.3cm}A5 & 8.44M & 1st & 128 & 13.24 & 12.58 & 18.30 & 19.77 & 32.98 & 17.77 \\
    A6 & 8.44M & 2nd & 128 & 13.53 & 14.04 & 16.92 & 18.98 & 33.48 & 17.69 \\
    A7 & 8.44M & 3rd & 128 & \textbf{12.76} & 12.77 & 16.50 & 20.61 & 32.72 & 17.24 \\
    A8 & 8.44M & 4th & 128 & 12.82 & 15.01 & \textbf{15.70} & 19.33 & 31.34 & 17.05 \\
    \bottomrule
  \end{tabular}
  }
\end{table*}

In Table \ref{tab:dat}, we investigate the effectiveness of using DAT with adapters. When adapters are used at the end of each encoder block, the overall DER performance improves from 17.66\% to 16.74\% with a bottleneck size of 32 (A2). This contributes to a relative improvement of 5.2\% over B2. The results also show that a larger bottleneck size does not further improve the model, and the best performance can be achieved by only 0.35M extra parameters. 

We also investigate the impact of using adapters only in one single encoder block instead of all the blocks. For fair comparison, we use a similar number of parameters to A2. When adapters are only used in the third (A7) and last (A8) encoder block with a bottleneck size of 128, the overall DER performance can be improved from 17.66\% to 17.24\% and 17.05\% when compared with B2, respectively. This shows that DAT with adapters is more effective in the later blocks. However, using adapters in all of the encoder blocks with a similar number of parameters still performs the best.

\begin{table}[htb]
  \caption{Summary and combination of our proposed methods.}
  \label{tab:summary}
  \centering
  \resizebox{\linewidth}{!}{
  \begin{tabular}{cM{1.5cm}cccccc}
    \toprule
    Setup & No. of Parameters & DH3 & VC & MD & AM & AS & Overall \\
    \midrule
    B1 & 8.10M & 13.48 & 17.18 & 19.74 & 21.36 & 34.15 & 19.24 \\
    B2 & 8.10M & 14.25 & 12.87 & 15.75 & 20.23 & 33.71 & 17.66 \\
    M6 & 8.17M & 14.19 & 12.62 & 14.33 & 18.87 & 33.12 & 16.99 \\
    A2 & 8.45M & \textbf{12.94} & \textbf{11.73} & 16.51 & 19.10 & \textbf{30.99} & 16.74 \\
    A2 + M6 & 8.52M & 13.15 & 12.17 & \textbf{14.19} & \textbf{18.60} & 33.83 & \textbf{16.59} \\
    \bottomrule
  \end{tabular}}
\end{table}

In Table \ref{tab:summary}, we summarize the best setups for our proposed methods along with the baseline models. When combining the best setups from DAT (A2) and MTL (M6), we achieve an overall DER of 16.59\% with only 0.42M extra parameters. This demonstrates that both our proposed techniques are effective. The combined setup (A2 + M6) improves the overall DER from 19.24\% to 16.59\% when compared with B1, which contributes to an absolute improvement of 2.65\%. This also improves the overall DER from 17.66\% to 16.59\% when compared with B2, contributing to an absolute improvement of 1.07\%.

\begin{table}[htb]
  \caption{Comparison of our proposed methods to other EEND-based published results with all number of speakers (ranging from 1 to 21).}
  \label{tab:comparison}
  \centering
  \resizebox{\linewidth}{!}{
  \begin{tabular}{lccccc}
    \toprule
    Model & DH3 & VC & MD & AM & AS \\ 
    \midrule
    SA-EEND \cite{eend_eda} & 22.64 & - & - & 27.70 & - \\ 
    EEND-EDA \cite{eend_eda} & 20.69 & - & - & 21.56 & - \\ 
    EEND-EDA AAM \cite{eend_eda_aam} & 19.99 & - & - & - & - \\ 
    \hdashline
    \rule{0pt}{0.3cm}EEND-EDA AAM (B2) & 20.15 & 26.21 & 15.75 & 20.23 & 33.71 \\ 
    + DAT (A2) & 19.28 & \textbf{26.06} & 16.51 & 19.10 & \textbf{30.99} \\ 
    + MTL (M6) & 20.34 & 26.52 & 14.33 & 18.87 & 33.12 \\ 
    + DAT \& MTL (A2 + M6) & \textbf{18.90} & 26.22 & \textbf{14.19} & \textbf{18.60} & 33.83 \\ 
    \bottomrule
  \end{tabular}}
\end{table}

In Table \ref{tab:comparison}, we compare our proposed methods with other EEND-based published results with all number of speakers. This demonstrates our proposed methods are capable of improving the overall DER performance on top of a strong and competitive EEND-based baseline (B2). For example, we improve DH3 over B2 from 20.15\% to 18.90\% using our best setup (A2 + M6).

\section{Analysis}
\label{analysis}

In this section, we study the effect of using adapters from the correct and incorrect domains, model generalizability to unseen domains, and effect of bypassing conversational summary vector in adapters.

\subsection{Use of adapters from correct/incorrect domains}
\begin{table}[htb]
  \caption{Effect of using adapters from different domains during inference for different datasets in our best setup (A2 + M6). When no adapter is used, the domain is denoted as `(No adapter)'.}
  \label{tab:domains}
  \centering
  \resizebox{\linewidth}{!}{
  \begin{tabular}{cccccccc}
    \toprule
    Domain/Dataset & DH3 & VC & MD & AM & AS & AF & AN \\
    \midrule
    DH3 & \textbf{13.15} & 23.63 & 27.93 & 37.73 & 42.59 & 29.81 & 29.46 \\
    VC & 22.46 & 12.17 & 22.53 & 33.79 & 40.27 & 25.47 & 20.69 \\
    MD & 23.35 & 21.36 & \textbf{14.19} & 59.23 & 58.79 & 40.54 & 39.14 \\
    AM & 27.10 & 13.65 & 25.13 & \textbf{18.60} & 35.98 & 25.27 & 19.72 \\
    AS & 28.20 & 14.63 & 24.97 & 21.40 & \textbf{33.83} & 25.37 & 20.24 \\
    (No adapter) & 20.76 & \textbf{11.58} & 21.35 & 25.70 & 34.66 & \textbf{23.09} & \textbf{18.76} \\
    \bottomrule
  \end{tabular}}
\end{table}

Table \ref{tab:domains} shows that the DER performance is usually significantly worse when adapters from the incorrect domains are used. In general, the best DER performance for any previously seen domain can be achieved by using the adapter from the correct (ground-truth) domain. This is true for all five seen domains except for VC. Since VC is collected from the YouTube channel, we hypothesize this can be due to the diverse data characteristics present in this dataset. In such case, domain adaptive training with adapters is still competitive but less effective.

\begin{table}[htb]
  \caption{Effect of whether to use adapters during inference. When adapters are used for seen domains, adapters from the ground-truth domains are used. No adapter is used for unseen domains (AF/AN).}
  \label{tab:adapters}
  \centering
  \resizebox{\linewidth}{!}{
  \begin{tabular}{cM{1.5cm}M{1.2cm}M{1.2cm}M{1.2cm}M{1.5cm}}
    \toprule
    Setup & No. of Parameters & Adapters for seen? & Overall (seen) & AF (unseen) & AN (unseen) \\
    \midrule
    B2 & 8.10M & - & 17.66 & 39.91 & 25.32 \\
    A2 + M6 & 8.52M & No & 21.96 & \textbf{23.09}  & \textbf{18.76} \\
    A2 + M6 & 8.52M & Yes & \textbf{16.59}  & \textbf{23.09}  & \textbf{18.76} \\
    \bottomrule
  \end{tabular}}
\end{table}

\subsection{Model generalizability to unseen domains}
\label{ssec:generalizbility}

We demonstrate our model achieves a stronger generalizability to unseen domains over the baseline. For unseen domains, all adapters are removed during inference because adapters from the ground-truth domains are not available. As shown in Table \ref{tab:adapters}, this achieves the best DER performance of 23.09\% and 18.76\% for unseen domains AF and AN, respectively. These results are significantly better than 39.91\% and 25.32\% in B2. This shows our model is capable of learning better domain-agnostic components than the baseline. Different from \cite{ali}, we do not use any data from AF and AN during training. However, our model still achieves a better DER performance than the AF baseline in \cite{ali}, which is 24.95\%.

When prior knowledge on data nature is available, adapters from AM and AS domains can also be used because of their similarity in data characteristics, i.e. meeting recordings. In Table \ref{tab:domains}, this gives a DER performance ranging from 25.27\% to 25.37\% and 19.72\% to 20.24\% for AF and AN respectively. These results correspond to the second and the third best DER. However, removing adapters during inference for unseen domains still perform the best.

\begin{table}[htb]
  \caption{Effect of using different adapter dropout values during training. Adapter dropout is not used during inference. The adapter from the ground-truth domain is used for each seen domain during inference. No adapter is used for unseen domains (AF/AN).}
  \label{tab:adapter_dropout}
  \centering
  \resizebox{\linewidth}{!}{
  \begin{tabular}{cM{1.5cm}M{1.2cm}M{1.2cm}M{1.2cm}M{1.5cm}}
    \toprule
    Setup & No. of Parameters &  Adapter dropout & Overall (seen) & AF (unseen) & AN (unseen) \\
    \midrule
    A2 + M6 & 8.52M & 0.0 & \textbf{16.59}  & \textbf{23.09}  & \textbf{18.76} \\
    A2 + M6 & 8.52M & 0.025 & 17.11  & 23.76  & 20.15 \\
    A2 + M6 & 8.52M & 0.05 & 17.76  & 29.98  & 21.07 \\
    A2 + M6 & 8.52M & 0.1 & 17.38  & 33.97  & 20.69 \\
    A2 + M6 & 8.52M & 0.2 & 17.99  & 34.74  & 24.14 \\
    \bottomrule
  \end{tabular}}
\end{table}

When adapters are removed during inference, this introduces training and testing mismatch. In order to reduce such mismatch, we study the effect of using different adapter dropout values during training. Adapter dropout has an effect of removing adapters during training based on a certain probability. Adapter dropout is not used during inference. In Table \ref{tab:adapter_dropout}, the results show that adapter dropout degrades the DER performance. We hypothesize that introducing adapter dropout can also confuse the learning of domain-agnostic components in the model. This matches to our expectation that adapters are more responsible to learn the domain characteristics in comparison to those domain-agnostic components.

\begin{table}[htb]
  \caption{Effect of bypassing conversational summary vector in adapters in our best setup (A2 + M6).}
  \label{tab:skip}
  \centering
  \resizebox{\linewidth}{!}{
  \begin{tabular}{cM{1.5cm}ccccccc}
    \toprule
    Setup & No. of Parameters & Bypass & DH3 & VC & MD & AM & AS & Overall \\
    \midrule
    A2 + M6 & 8.52M & No & \textbf{13.15} & 12.17 & \textbf{14.19} & \textbf{18.60} & 33.83 & \textbf{16.59} \\
    A2 + M6 & 8.52M & Yes & 13.60 & \textbf{12.12} & 17.53 & 20.10 & \textbf{30.58} & 17.36 \\
    \bottomrule
  \end{tabular}}
\end{table}

\subsection{Conversational summary vector bypass in adapters}
\label{ssec:bypass}

We conduct an ablation study whereby the conversational summary vector bypasses adapters. In Table \ref{tab:skip}, the results show that the overall DER performance degrades from 16.59\% to 17.36\%. This suggests that adapting the conversational summary representation with adapters makes the auxiliary domain classification task easier, which improves the diarization performance.

\section{Conclusion}

In this paper, we demonstrated the effectiveness of leveraging domain information to train a single end-to-end diarization model for multiple domains. We proposed the use of domain adaptive training with adapters, and multi-task learning with an auxiliary domain classification task. For domains seen during training, the combination of our methods improved the absolute DER performance from 19.24\% to 16.59\% when compared with the domain-specific baseline. We also improved the DER from 17.66\% to 16.59\% when compared with the multi-domain baseline. This contributes to a significant relative improvement of 13.8\% and 6.1\% respectively. Importantly, we have demonstrated that our model achieves a stronger generalizability towards previously unseen domains during inference when adapters are removed. When comparing to the multi-domain baseline, this improved the absolute DER from 39.91\% to 23.09\% and 25.32\% to 18.76\% respectively for two unseen domains.

\bibliographystyle{IEEEbib}
\bibliography{refs}

\end{document}